\documentstyle[prl,aps,preprint]{revtex}    
\begin{document}
\draft
\tightenlines
\title{Lifetime of collective isospin
rotations of a quantum meson field}
\author{Y. Tsue\footnote{
Permanent address : Department of Material Science, Kochi
University, Kochi 780-8520, Japan}
and D.~Vautherin}
\address{L.P.N.H.E., L.P.T.P.E., Universit\'es Paris VI/VII,
F-75252, Paris Cedex 05, France}
\author{T.~Matsui} \address{Institute of Physics, University of Tokyo,
Komaba, Meguro-ku, Tokyo 153, Japan} 

\date{\today} 

\maketitle

\begin{abstract}
We calculate the lifetime of the collective isospin rotating solutions
which have been found recently in the case a quantum N-component meson field
with exact O(N) symmetry. For this purpose we take into account the small
breaking of the O(N) symmetry associated to the non vanishing mass of the
pion. This term induces a coupling between collective rotations and intrinsic
meson excitations. We evaluate the associated damping time in the 
framework of
linear response theory. We find damping times of the order of 100 fm/c,
i.e. substantially longer than reaction times.
\end{abstract}

\narrowtext

\newpage

%
\def\mib#1{\mbox{\boldmath $#1$}}
\section{Introduction}
The forthcoming generation of ultrarelativistic nuclear
colliders at Brookhaven and CERN opens the
perspective of exploring the dynamics of matter at very high energy
density. Among the most central issues in this field is the production and
study of the quark-gluon plasma. Another attractive question is the physics
of high density meson fields out equilibrium which could evolve on a
time scale of the order of a fm/$c$ or less. Such systems are of interest
for the understanding of the field theoretical problems involved in the first
stages of the early universe. They may also exhibit collective phenomena
closely connected to fundamental symmetry properties such as chiral
symmetry. The formation, the stability and the damping time of such 
collective
modes are important, yet unanswered questions, although some progress in 
this
direction has been carried out recently \cite{DEVEGA,SALGADO,SERREAU}.

Collective modes in a high density pion field have been studied in the
context of classical evolution equations by Anselm and Ryskin
\cite{ANSELM} and also by Blaizot and Krzywicki \cite{BLAIZOT} who
focused on the issue of Lorentz invariant boundary conditions.
These time-dependent solutions of classical meson fields attracted
wider attention from the point of view of non-equilibrium dynamics of
the chiral phase transition \cite{BJORKEN,RAJAGOPAL}.  Recently quantum
fluctuations in the description of these modes were introduced
\cite{PRD,TSUE}. Particular analytic solutions of the mean field
evolution equations corresponding to quantum rotations in isospin
space were constructed, both in the zero temperature case\cite{PRD} as
well as in the finite temperature case \cite{TSUE}.

In the limit of exact chiral symmetry these rotations decouple from
intrinsic meson excitations. However when an explicit symmetry
breaking term, which generates non zero pion mass, is introduced, then
this term will act as an anisotropic (in isospin space) external
potential for the isospin rotation which causes coupling of the
collective rotation and intrinsic mesonic excitations of the system
and this coupling in turn results in a dissipation of the collective
excitation energy to mesonic excitation energy.  This effect is very
similar to the Landau damping of collective oscillations in
collisionless plasma.  Indeed, as we shall see below, the symmetry
breaking term is viewed in the isospin rotating frame as a periodic
time-dependent external perturbation, the linear response of the
system to such an external oscillatory source contains collisionless
dissipation, a characteristic feature of the Landau damping
mechanism generalized to other systems. 
As a result the collective ``angular momentum" for the isospin rotation
diminishes. Since the collective isospin angular momentum may be
proportional to the ``moment of inertia" of the isospin rotation (the square
of the condensate) times the corresponding angular frequency, both the
angular frequency and the radius of the chiral circle, which are
related each other in the gap equation, may become small.

It is the purpose of the present work to evaluate the corresponding
damping time. In order to perform our calculations we will assume
that the coupling between collective rotations and meson excitations
via the symmetry breaking term is small. This is the case when the
kinetic energy density of collective rotations is large compared to
the change in the potential energy density induced by symmetry
breaking term. This can be seen to be the case when the frequency
$\omega$ of rotational motion is large compared to twice the pion
mass (see below).

In the following we will investigate properties of the damping time
such as its dependence on excitation energy density, angular frequency
and temperature. Another question of interest is to work out the
spectrum of emitted particles.  Experimental issues related to the
previous questions are the observability of classical meson field as
suggested first by Anselm and Ryskin \cite{ANSELM}.  In particular
it is obviously important to have a reliable estimate of the
lifetime of the collective modes of the pion field in order to be able
to check that it is longer than the reaction time. Also it is of
great interest to have estimates of the lifetimes of various modes
in order to find out whether damping could act as a filter to select
the most collective states as the nuclear collision proceeds. This
information is crucial to discuss the formation of collective
modes. 

In section 2 we review the isospin rotating solutions of the quantum
mean field equations which were obtained in reference \cite{TSUE},
hereafter referred to as I. These solutions correspond to static
solutions in the isospin rotating frame in the case of exact chiral
symmetry. 
In section 3 we consider the case where there is an
explicit symmetry breaking term. We show that this term generates a
periodic external source in the rotating frame and we construct an
explicit expression of the source. In section 4 we describe the
general formalism of linear response theory in the rotating frame
which is suited to deal with the case of a small symmetry breaking. As
a warm up exercise we deal in section 5 and 6 
with the simple case of the
bare response in which residual interactions are neglected. The
realistic situation where interactions are included is discussed in
section 7. The expression of the damping time is derived in section
8 and a discussion of its physical content is presented in section
9.
In Appendix A, we investigate the pionic excitation in the 
chiral limit in the linearized mean field approach and 
show that our approach is consistent with the 
Goldstone theorem.

\section{Review of rotating solutions of the mean field equations}
Time-dependent solutions of the classical equations of motion of the
non-linear sigma model have been constructed and discussed by Anselm
and Ryskin\cite{ANSELM} and others\cite{BLAIZOT}. These solutions
describe field rotations in its internal symmetry space which may be
considered to appear as a result of the spontaneously broken internal
symmetry, similar to nuclear deformation and rotation.  Here we review
a quantum extension of the Anselm-Ryskin solutions in the framework of
the mean-field description as obtained in \cite{PRD} and in I.

For N-component scalar fields $\varphi_a(x)$ with a bare mass
$m_0^2$ and a coupling constant $\lambda$,
the evolution of the mean field ${\bar \varphi}$ is governed by\cite{PRD}
\begin{equation} \label{MEANFIELD}
\{ \Box+ m_0^2 + \frac{\lambda}{6} {\bar \varphi}^2 +
\frac{\lambda}{6} {\rm trace} S(x,x)+
\frac{\lambda}{3} S(x,x) \} {\bar \varphi}=0  \mbox{,}
\end{equation}
where the trace runs over the index $a$ and the N$\times$N matrix operator
$S$ is the Feynman propagator, $i S^{-1}= \Box+m^2(x)+ i \varepsilon$,
with the mass matrix given by
\begin{equation} \label{gap}
m^2(x)= m_0^2 +\frac{\lambda}{6} {\bar \varphi}^2(x) +
\frac{\lambda}{6} {\rm trace} S(x,x) 
+ \frac{\lambda}{3} {\bar \varphi}(x) \times {\bar \varphi}(x)
+ \frac{\lambda}{3} S(x,x) \mbox{,}
\end{equation}
where the symbol ${\bar \varphi}(x) \times {\bar \varphi}(x)$ denotes
the $N \times N$ matrix whose matrix $(a,b)$ element is ${\bar
\varphi}_a (x) \times {\bar \varphi}_b (x)$.  To construct solutions to the
above coupled equations, we adopt the following Ansatz:
\begin{equation} \label{ansatz}
{\bar \varphi}(x)= U(x) {\bar \varphi}_r=
\exp \{i(\omega t- {\mib q} \cdot {\mib x}) \tau_y \} {\bar \varphi}_r,
\end{equation}
and
\begin{displaymath}
S (x, y) = U(x) S_r (x,y) U^{\dagger}(y) \ ,
\end{displaymath}
where ${\bar \varphi}_r$ is the amplitude of the mean field in the
``rotating frame'', $\tau_y$, the second Pauli matrix, is the
generator of the rotation in the subspace of O(N),
\begin{equation}\label{4}
\tau_y =
\pmatrix{
0 & -i & 0 & \cdots \cr
i & 0  & 0 & \cdots \cr
0 & 0  & 0 & \cdots \cr
\vdots & \vdots & \vdots & \ddots
}
\ ,
\end{equation}
and $\omega$ and
${\mib q}$ are arbitrary parameters to characterize space-time
dependence of the rotation.  The propagator $S_r$ in the
rotating frame takes the form of
\begin{equation} \label{PROPAGATOR}
S_r(x,y)= \frac{i}
{(\partial_{\mu} + i q_{\mu} \tau_y)(\partial^{\mu} + i q^{\mu} \tau_y)
+M^2+ i \varepsilon},
\end{equation}
where
\begin{equation} \label{MSQUARE}
M^2 = U (x) m^2 (x) U^{\dagger} (x).
\end{equation}
In the rotating frame, the condensate $\bar \varphi_r $ and the mass
matrix $M^2$ satisfy the mean field equation (\ref{MEANFIELD}) and
the gap equation (\ref{gap}) with $S (x, x)$ and $\bar \varphi
(x)$ replaced by $S_r (x,x)$ and ${\bar \varphi}_r$, respectively,
and one can show that $M^2$ becomes independent of space and time.
These two equations form a closed set to determine the values of $M$
and $\bar \varphi_r$ self-consistently. 

In the limit $q$=0 one recovers the self-consistent equations
specifying the properties of the vacuum state, namely the associated values of
the mass matrix $m$ and of the condensate $\varphi_0$. These equations are
reproduced in Appendix A (equations (\ref{pion14}) - (\ref{pion17})). 
The mass
matrix $m$ is diagonal in the vacuum state, with $N-1$ meson masses being
equal to a value $\mu$ (which corresponds in the sigma model $N=4$ to the pion
mass), the last one being equal to $M$ (the sigma meson mass in the sigma
model). It is worthwhile noting that even in the chiral limit (no symmetry
breaking term) the pion mass in the mean field approximation does not vanish
(see Appendix A). However the existence of rotating solutions guarantees that
the linearized mean-field evolution equations do have solutions with zero
frequencies. As a result the pion masses in this more elaborate framework do
vanish as shown by an explicit calculation described in Appendix A. This
implies that the Goldstone theorem is recovered only at the level of
the linearized mean-field evolution equations. This is in agreement with the
discussions already given by other authors 
\cite{DMITRASINOVIS,OKOPINSKA,NAUS,AOUISSAT}. 

The solutions we have just reviewed concern only the zero temperature
case, i.e. pure states. Generalization of these solutions to finite
temperature  was reported in I. In this case one needs to consider
statistical mixtures. These are most conveniently handled by
introducing the corresponding density operator formalism
\cite{EBOLI,BALIAN}. In this formalism the mean field approximation
is formulated simply by stating that the density operator $D$
describing the system is given at each time $t$ by the exponential of
a quadratic form in the field operators. In this case the
expectation value of an arbitrary operator
\begin{equation} \label{meanvalue}
\langle {\bf O} \rangle = {\rm Trace} \{ D  {\bf O} \}
\end{equation}
can be
expressed in terms of the generalized density matrix ${\cal M}$ defined by
\begin{equation}\label{15}
{\cal M}({\mib x},{\mib y};t) + \frac{1}{2} =
\pmatrix{
-i \langle {\hat \varphi}_a({\mib x}) {\hat \pi}_b({\mib y}) \rangle
& \langle {\hat \varphi}_a({\mib x}) {\hat \varphi}_b({\mib y}) \rangle  
\cr
\langle {\hat \pi}_a({\mib x}) {\hat \pi}_b({\mib y}) \rangle
&i \langle {\hat \pi}_a({\mib x}) {\hat \varphi}_b({\mib y}) \rangle  }
\ .
\end{equation}
In this equation $\pi_b({\mib x})$ is the momentum conjugate
to  $\varphi_b({\mib x})$. We have used the notation
\begin{equation} \label{hatphi}
{\hat \varphi}_a({\mib x}) =  \varphi_a({\mib x})-
 \langle \varphi_a({\mib x}) \rangle.
\end{equation}
Expectation values of field operators are defined by equation
(\ref{meanvalue}).
The mean field evolution equation for the generalized density matrix is
obtained in this formalism by stating that the trace of the Liouville 
equation
for $D$ multiplied by an arbitrary quadratic form in the field operators
vanishes, i.e.
\begin{equation}\label{liouville}
{\rm Trace}\ \{(i {\dot D} - [H,D]) {\hat \varphi}_a({\mib x})
{\hat \varphi}_b({\mib y}) \}=0,
\end{equation}
where $H$ is the hamiltonian of the O(N) model whose density reads
$$
{\cal H}_{\rm O(N)}({\mib x})
=\frac{1}{2}\pi_a({\mib x})\pi_a({\mib x})
+\frac{1}{2}\nabla\varphi_a({\mib x})\nabla\varphi_a({\mib x}) 
+\frac{1}{2}m_0^2 \varphi_a({\mib x})\varphi_a({\mib x})
+\frac{\lambda}{24}(\varphi_a({\mib x})\varphi_a({\mib x}))^2 
\ .
$$
By calculating the commutators in this formula one obtains the following
evolution equation for the generalized density matrix
\begin{equation}\label{hfb}
i\dot{\cal M}(t) = [ {\cal H} \ , \ {\cal M}(t) ] \ ,
\end{equation}
where
\begin{eqnarray}
{\cal H}&=&\pmatrix{ 0 & 1 \cr
                   \Gamma & 0
                 } \ , \label{staticcalh}\\
\Gamma^{ab}&=& \left(- \Delta
+ m_0^2+\frac{\lambda}{6}{\bar \varphi}^2
                      +\frac{\lambda}{6}
 \langle {\hat \varphi}_c{\hat \varphi}_c \rangle \right) \delta_{ab}
+\frac{\lambda}{3}({\bar \varphi}_a{\bar \varphi}_b
+ \langle {\hat \varphi}_a{\hat \varphi}_b \rangle ) \ , 
\label{staticgamma}
\end{eqnarray}
and $1$ is an $N\times N$ unit matrix in the isospin space.
The detailed derivation of this equation is given in I.

Static solutions of these equations corresponding to thermal equilibrium
at a given temperature $T$ are obtained by writing that the generalized
density matrix ${\cal M}$ commutes with the mean field hamiltonian ${\cal
  H}$. In this case one finds a Boltzmann-Gibbs type distribution for the
density operator
\begin{equation}\label{hotstaticD}
D=\frac{1}{Z} e^{-\beta W} \ , \qquad
Z={\rm Trace}\ e^{-\beta W} \ ,
\end{equation}
where $\beta=1/k_{\rm B}T$. In the previous equation $W$ is
the Hartree-Bogoliubov Hamiltonian
\begin{eqnarray}\label{hfbhamiltonian}
W&=&\int d^3{\mib x} \biggl\{
 \frac{1}{2}{\hat \pi}_a({\mib x}){\hat \pi}_a({\mib x})
 + \frac{1}{2}\nabla{\hat \varphi}_a({\mib x})\nabla{\hat \varphi}_a
({\mib x})
 + \frac{m_0^2}{2}{\hat \varphi}_a({\mib x}){\hat \varphi}_a({\mib x})
\nonumber\\
 & & +\frac{\lambda}{12}\langle \varphi_a({\mib x})\varphi_a({\mib x})\rangle
   {\hat \varphi}_b({\mib x}){\hat \varphi}_b({\mib x}) 
 + \frac{\lambda}{6}\langle \varphi_a({\mib x})\varphi_b({\mib x})
\rangle
   {\hat \varphi}_a({\mib x}){\hat \varphi}_b({\mib x}) \biggl\} \ .
\end{eqnarray}

In order to build rotating solutions at finite temperature it is
convenient to write the equation of motion for the generalized
density matrix in an isospin rotating frame characterized by a
frequency $\omega$ and a wave vector ${\mib q}$ as defined by
equation (\ref{ansatz}). 
This is done again by gauge transforming the density
matrix:
\begin{equation}
{\cal M}({\mib x},{\mib y};t)  = U ({\mib x},t) {\cal M}_r
({\mib x},{\mib y};t)
U^{\dagger}({\mib y},t) \ .
\end{equation}
Then the Liouville equation is transformed to
\begin{equation} \label{rotatingHFB}
i\dot{\cal M}_r = [ {\cal H}_r \ , \ {\cal M}_r ] ,
\end{equation}
where the mean field hamiltonian in the isospin rotating frame is given by
\begin{eqnarray}
{\cal H}_r & = &
U^{\dagger} (x) ( - i \frac{\partial}{\partial t} + {\cal H} ) U (x) =
\pmatrix{ \omega \tau_y & 1 \cr
                   \Gamma_r & \omega\tau_y
                 } \ , \label{rotHAMILTONIAN}\\
\Gamma^{ab}_r &=& \left(-(\nabla-i{\mib q}\tau_y)^2
+ m_0^2+\frac{\lambda}{6}{\bar \varphi_r}^2
                      +\frac{\lambda}{6}
 \langle {\hat \varphi}_c{\hat \varphi}_c \rangle_r \right) \delta_{ab}
+\frac{\lambda}{3}({\bar \varphi}_{r,a}{\bar \varphi}_{r,b}
+ \langle {\hat \varphi}_a{\hat \varphi}_b \rangle_r ) \ . \label{rotGAMMA}
\end{eqnarray}
In the above expression, the subscript $r$ indicates that the variable
is taken in the isospin rotating frame. In the following, we omit these
subscripts unless it is very confusing.

We construct the isospin rotating solutions at finite temperature
by imposing that the density matrix takes a static, thermally equilibrated
form in the rotating frame.
In this the case ${\cal H}_r $ and ${\cal M}_r$ commute
and therefore they can be diagonalized simultaneously:
\begin{equation}\label{26}
{\cal H}_0 \pmatrix{ u^l \cr v^l } = E^l \pmatrix{ u^l \cr v^l } \ , 
\quad
{\cal M}_0 \pmatrix{ u^l \cr v^l } = f^l \pmatrix{ u^l \cr v^l } \ . 
\end{equation}
Here, the static and thermally equilibrated case is described
with the subscript 0.  As is
expected, $(u^*, -v^*)$ is also an eigenvector with an negative eigenvalue
$-E$ and $-f$.
Here let us introduce a diagonal basis set $\{ |a, {\mib k}, \tau \rangle 
\}$
for ${\cal H}_0$ and ${\cal M}_0$ in momentum space,
where $a$ and ${\mib k}$ represent isospin index and momentum, 
respectively,
and $\tau = +$ or 1 ($-$ or 2) represents the state with a positive
(negative) eigenvalue. Then we obtain
\begin{eqnarray}\label{27}
& &{\hat {\cal H}}_0 |a,{\mib k},\tau \rangle
= E_a^{\tau}({\mib k}) |a,{\mib k},\tau \rangle \ ; \ 
E_a^{\pm}({\mib k}) = \pm E_a({\mib k}) \ , \nonumber\\
& &{\hat {\cal M}}_0 |a,{\mib k},\tau \rangle
= f_a^{\tau}({\mib k}) |a,{\mib k},\tau \rangle \ ;  \ 
f_a^{\pm}({\mib k}) = \pm\left(n_a({\mib k})+\frac{1}{2}\right) , 
\end{eqnarray}
where $E_a({\mib k})$ can be directly calculated from
${\cal H}_0$ and is given in I and 
$n_a({\mib k})=1/(\exp(\beta E_a({\mib k})) -1)$ is a bose distribution
function.

The density operator which gives a thermal distribution in
an isospin rotating frame can be shown to be of the form
\begin{equation}\label{hotrotatingD}
D(\omega, {\mib q}) =\frac{1}{Z} e^{-\beta {\overline W}
(\omega, {\bf q})
} \ , \qquad
Z={\rm Trace}\ e^{-\beta {\overline W}} \ ,
\end{equation}
which is the same as before except for the presence of a modified
Hartree-Bogoliubov Hamiltonian
\begin{equation}\label{modifHFB}
{\overline W}(\omega, {\mib q})=
W -\omega I_{ab} + {\mib q}\cdot {\mib I}_{ab}
+\frac{{\mib q}^2}{2}
\int d^3{\mib x} {\hat \varphi}_c({\mib x}){\hat \varphi}_c({\mib x})
\ .
\end{equation}
In this equation the operator $I$ is the generator of isospin rotations
\begin{eqnarray}\label{14}
I_{ab}&=&\int d^3{\mib x} [{\hat \varphi}_a({\mib x}){\hat \pi}_b
({\mib x})
                      -{\hat \varphi}_b({\mib x}){\hat \pi}_a({\mib x}) ] 
\ ,
\nonumber\\
{\mib I}_{ab}&=&\int d^3{\mib x} [{\hat \varphi}_a({\mib x})\nabla
{\hat \varphi}_b({\mib x})
                      -{\hat \varphi}_b({\mib x})\nabla
{\hat \varphi}_a({\mib x}) ] \ .
\end{eqnarray}
We choose $a=1$ and $b=2$ for the rotation in the isospin space
in accord with the ansatz (\ref{ansatz}).
In what follows we will denote by ${\cal M}_0(\omega, {\mib q},T)$
the associated generalized density matrix.

\section{Classical isospin rotations with small symmetry breaking}

In this section we consider the case where a small explicit symmetry
breaking term is added to the Hamiltonian density, i.e.
\begin{equation} \label{brokenhdens}
{\cal H}({\mib x})=
{\cal H}_{\rm O(N)}({\mib x})
-c\varphi_a({\mib x})\delta_{a1} \ .
\end{equation}
We wish to work out the effect of the second term on our rotating solutions
described by ${\cal M}_0(\omega, {\mib q},T)$ in the rotating frame. In a 
first
step let us investigate the change it produces on the isospin rotating
solutions obtained in the context of classical dynamics by Anselm and 
Ryskin
\cite{ANSELM}.
The classical equation of motion is derived as
\begin{equation}\label{2}
\left(\partial_\mu \partial^\mu +
m_0^2 + \frac{\lambda}{6} {\bar \varphi}^2 \right){\bar \varphi} 
= c {\mib 1}
\ .
\end{equation}
We are interested in the isospin rotating solution.
In rotating frame, the expectation value (condensate) of the field is
written as
\begin{equation}\label{3}
{\bar \varphi} = e^{iqx\tau_y}({\bar \varphi}^{(0)} + 
\delta \varphi (x) )
\end{equation}
with $q=(\omega, {\mib q})$.
Here we introduce the condensate $\varphi_0$, which does not depend on 
space
and time, pointing in the direction of the first isospin :
\begin{equation}\label{5}
{\bar \varphi}^{(0)} =
 \pmatrix{ \varphi_0 \cr
           0 \cr
           \vdots \cr
           0 } \ .
\end{equation}
The condensate $\varphi_0$ satisfies a gap equation without $c$ in the 
chiral
symmetry broken phase :
\begin{equation}\label{6}
-q^2 + m_0^2 + \frac{\lambda}{6} \varphi_0^2 = 0 \ .
\end{equation}
The quantity $\delta\varphi (x)$ appears due to non-vanishing $c$, that is,
the existence of the explicit chiral symmetry breaking term.
Then the equation of motion for the condensate is reduced to the following
equation of motion for $\delta \varphi(x)$ :
\begin{equation}\label{7}
\biggl\{ [ (\partial_\mu +iq_\mu \tau_y)(\partial^\mu + iq^\mu 
\tau_y)]_
{ij}
+(m_0^2 +\frac{\lambda}{6}\varphi_0^2)\delta_{ij} 
+\frac{\lambda}{3}{\bar \varphi}_i^{(0)} {\bar \varphi}_j^{(0)} 
\biggl\} \delta\varphi_j (x) =
c
\pmatrix{
\cos qx \cr
\sin qx \cr
0 \cr
\vdots \cr
0\cr
}_i \ .
\end{equation}
This equation can be solved up to the order of $c$ :
\begin{eqnarray}
\delta \varphi (x) &=&
\frac{c}{A}\pmatrix{ (m_\pi^2-4q^2) \cos qx \cr
                     (m_\sigma^2-4q^2) \sin qx \cr
                     0 \cr
                     \vdots \cr
                     0
                   } \ , \label{8}\\
A&=&m_\sigma^2 m_\pi^2 -2q^2(m_\sigma^2 + m_\pi^2) \ , \label{9}
\end{eqnarray}
where $m_\pi^2 = m_0^2+(\lambda/6)\varphi_0^2$ and
$m_\sigma^2 = m_0^2+(\lambda/2)\varphi_0^2$.
$\delta \varphi (x)$ depends on both time and space in the isospin rotating
frame. This solution produces a term which plays a role of an external source
for collective isospin rotation of quantum meson fields.

\section{Linearized mean-field dynamics in the isospin rotating frame}

In this section we examine the effect of the symmetry breaking term on
collective isospin rotations in the mean field approximation. For this 
purpose
we consider mean field dynamics in the presence of the source term 
derived in
the previous section. The boundary condition we consider is the following. 
We
assume that at time $t$=0 a collective rotating state has been created, 
i.e., we
take the density matrix in the rotating frame at time $t$=0
to be  ${\cal M}_0(\omega, {\mib q},T)$. We also assume that the symmetry
breaking term is small enough to allow the use of linear response
theory. Writing the generalized density matrix ${\cal M}(t)$ as
\begin{equation}
{\cal M}(t)={\cal M}_0 + \delta {\cal M}(t) \ , \label{19}
\end{equation}
and retaining only first order terms in $\delta {\cal M}$ in the evolution
equation, one finds
\begin{equation}\label{21}
i\delta{\dot{\cal M}}= [{\cal H}_0 \ , \delta {\cal M}]
+[\delta {\cal H} \ , {\cal M}_0 ]
+[\delta {\cal H}_{\rm ext} \ , {\cal M}_0] \ ,
\end{equation}
where
\begin{equation}\label{22}
    \delta{\cal H} = \pmatrix{ 0 & 0 \cr
                              \delta\Gamma & 0
                         } \ , \
    \delta{\cal H}_{\rm ext} = \pmatrix{ 0 & 0 \cr
                                \delta\Gamma_{\rm ext} & 0
                         }
\end{equation}
and
\begin{eqnarray}\label{23}
   \delta\Gamma^{ab}&=&
\frac{\lambda}{6} \left(\delta \langle {\hat \varphi}_c{\hat \varphi}_c
  \rangle\right) \delta_{ab}
+\frac{\lambda}{3} \delta\langle {\hat \varphi}_a{\hat \varphi}_b \rangle
\ , \nonumber\\        
   \delta\Gamma_{\rm ext}^{ab}&=&
\frac{\lambda}{3} \varphi_0 \delta\varphi_1 \delta_{ab}
+\frac{\lambda}{3} ({\bar \varphi}_a^{(0)} \delta\varphi_b
+\delta\varphi_a {\bar \varphi}_b^{(0)} ) \nonumber\\
&=&X_{ab}^{(-)}(q) e^{i{\bf q}\cdot{\bf x}} e^{-i\omega t} +
X_{ab}^{(+)}(q) e^{-i{\bf q}\cdot{\bf x}} e^{i\omega t} \ ,
\end{eqnarray}
while
\begin{equation}\label{24}
X_{ab}^{(\mp)}(q)=\frac{c\lambda \varphi_0}{6A}
\biggl\{(m_{\pi}^2-4q^2)(2+\tau_z)_{ab}
\pm i(m_{\sigma}^2-4q^2)(\tau_x)_{ab}
\biggl\}
 \ .
\end{equation}
Here $\tau_x$ and $\tau_z$ are $N\times N$ matrices in the isospin space
defined as 
\begin{equation}\label{25}
\tau_x =
\pmatrix{
0 & 1 & 0 & \cdots \cr
1 & 0  & 0 & \cdots \cr
0 & 0  & 0 & \cdots \cr
\vdots & \vdots & \vdots & \ddots
} \ , \quad
\tau_z =
\pmatrix{
1 & 0 & 0 & \cdots \cr
0 & -1  & 0 & \cdots \cr
0 & 0  & 0 & \cdots \cr
\vdots & \vdots & \vdots & \ddots
}
\ .
\end{equation}
It should be noted that $\delta{\cal H}_{\rm ext}$ appears due to the shift
of the condensate (mean field) associated with non-vanishing $c$.
We regard this term depending on both time and space as a classical source
term associated with the existence of the explicit chiral symmetry breaking
term in the isospin rotating frame.
We will investigate the response for this source in the succeeding sections.
Also, $\delta{\cal H}$ appears because of the coupling of the intrinsic
meson excitations and the classical rotating mean field in the isospin 
space.

In the basis which diagonalizes both ${\cal H}_0$ and ${\cal M}_0$
the linearized equation for $\delta {\cal M}$
is simply expressed as
\begin{eqnarray}\label{28}
i\langle a,{\mib k},\tau | \delta{\dot {\cal M}} |b, {\mib k}',\tau' 
\rangle
&=&
(E_a^{\tau}({\mib k})-E_b^{\tau'}({\mib k}'))
\langle a,{\mib k},\tau | \delta{\cal M} |b, {\mib k}',\tau' \rangle
\nonumber\\
& &+
(f_b^{\tau'}({\mib k}')-f_a^{\tau}({\mib k}))
\langle a,{\mib k},\tau | \delta{\cal H}+\delta {\cal H}_{\rm ext}
|b, {\mib k}',\tau' \rangle \ .
\end{eqnarray}
Here the relation between original matrix elements in (\ref{22})
and the new matrix elements in the diagonal basis is given as
\begin{equation}\label{29}
\langle a,{\mib k},\tau | \delta{\cal H}_{\rm (ext)}
|b, {\mib k}',\tau' \rangle 
=\left[U^{-1}({\mib k})\cdot \pmatrix{0 & 0 \cr
                       \delta\Gamma_{\rm (ext)}({\mib k},{\mib k}') & 0}
\cdot U({\mib k}') \right]_{a\tau, b\tau'} \ ,
\end{equation}
where
\begin{equation}\label{30}
U({\mib k})_{a\tau,l\tau'}=\pmatrix{u_a^l({\mib k}) & u_a^{l*}({\mib k})
\cr
              v_a^l({\mib k}) & -v_a^{l*}({\mib k}) }_{\tau\tau'}  
\ .
\end{equation}

\section{Construction of the bare response}

A usual step in the calculation of the full response function 
to an external
field is the construction of the so-called bare response function. This
function corresponds to the case of a simplified evolution equation in 
which
the mean field hamiltonian ${\cal H}(t)$ is approximated at each time $t$ 
by
the equilibrium mean-field hamiltonian  ${\cal H}_0$, which gives
\begin{equation} \label{BARE-EVOL}
i {\dot{\cal M}}= [{\cal H}_0 \ , {\cal M}].
\end{equation}
For this evolution equation it is straightforward to calculate the 
response
of the system (assumed to be in the state ${\cal M}_0$ at time 
$t=-\infty$) to a small external field of the form
\begin{equation} \label{BARE-EXTERNAL}
   \delta\Gamma_{\rm ext}^{ab}=
X_{ab} e^{i{\bf q}\cdot{\bf x}} e^{-i\omega t},
\end{equation}
where the frequency $\omega$ is assumed to contain a vanishingly
small imaginary part to guarantee an adiabatic switching of the
field. Because of translational invariance in time and space,
the change in the density matrix at time $t$ is of the form
\begin{equation} \label{DENSITY-CHANGE}
\delta\langle{\hat \varphi}_a({\mib x}){\hat \varphi}_b({\mib x})\rangle=
Y_{ab} e^{i{\bf q}\cdot{\bf x}} e^{-i\omega t}.
\end{equation}
When the evolution preserves isospin, then the matrix $Y$ is also
proportional to $X$.
The bare response function
is defined as the limit of the proportionality
coefficient when the external field strength tends to zero
\begin{equation} \label{BARE-RESPONSE}
\Pi^{(0)}_{ab} (\omega, {\mib q})= \lim_{X \to 0} Y/X.
\end{equation}
In this formula we have made explicit the dependence of the bare response
function $\Pi^{(0)}$ on the
frequency $\omega$, on the momentum ${\mib q}$ 
and on the isospin indices $a$ and
$b$.

To obtain an explicit form of the bare response we take matrix elements of 
the
evolution equation in the basis which diagonalizes both ${\cal H}_0$ and
${\cal M}_0$. We obtain for the changes in the elements of the density 
matrix
\begin{equation} \label{DELTA-MZERO}
\langle a,{\mib k},\tau | \delta{\cal M} |b, {\mib k}',\tau' \rangle
=\frac{f_b^{\tau'}({\mib k}')-f_a^\tau({\mib k})}
 {\omega -(E_a^\tau({\mib k})-E_b^{\tau'}({\mib k}'))}
\langle a,{\mib k},\tau | \delta{\cal H}_{\rm ext}
|b, {\mib k}',\tau' \rangle ,
\end{equation}
where the matrix elements of the external hamiltonian read
\begin{equation}\label{HEXT-MAT}
\langle a,{\mib k},\tau | \delta{\cal H}_{\rm ext}
|b, {\mib k}',\tau' \rangle 
=
X_{lm} \left[
\pmatrix{u_l^{a*}({\mib k}) \cr -u_l^a({\mib k}) }
\pmatrix{ u_m^{b}({\mib k}') , u_m^{b*}({\mib k}') }
\right]_{\tau\tau'} 
\delta({\mib k}'-{\mib k} + {\mib q})e^{ -i\omega t} \ .
\end{equation}
The evaluation of these elements requires the knowledge of the mode 
functions
$u_a^{l}({\mib k})$, defined by the eigenvalue equation (\ref{26}) and
obtained in analytical form in I. 
Also, we can derive $v_a^l({\mib k})$ by using of 
$u_a^l({\mib k})$ and (\ref{26}) : 
$v_a^l({\mib k})=E_a({\mib k})u_a^l({\mib k})
-\omega \sum_b(\tau_y)_{ab} u_b^l({\mib k})$.
For more simplicity 
we use in what follows the approximate form which corresponds to the 
limit
$q=0$ :
\begin{equation}\label{MODE-FUNCTIONS}
u_a^l ({\mib k}) \approx -\delta_{al} \frac{1}{\sqrt{2E_a({\mib k})}} 
\ ,
\quad
E_a({\mib k})=\sqrt{{\mib k}^2+M_a^2} \ .
\end{equation}
We also use the approximate values of the masses corresponding to
$q=0$, namely, $M_1=m_{\sigma}$ and
$M_2=M_3=\cdots =M_N =m_{\pi}$, when we estimate the damping
time of the collective isospin rotation.

By combining these equations we find that the change in the density
distribution reads
\begin{eqnarray} \label{PHI-PHI-CHANGE}
\delta\langle{\hat \varphi}_a({\mib x}){\hat \varphi}_b({\mib x})
\rangle
&=& X_{lm}\ e^{ i{\bf q}\cdot{\bf x}}e^{- i\omega t}  
\int\frac{d^3{\mib k}}{(2\pi)^3} \ 
\frac{n_a({\mib k})+n_b({\mib k} - {\mib q})+1}
{\omega-(E_a({\mib k})+E_b({\mib k} - {\mib q}))} \nonumber\\
& &\qquad\qquad\qquad\qquad\qquad\times
 u_a^{l'}({\mib k})u_l^{l'(*)}({\mib k})
u_m^{m'(*)}({\mib k} - {\mib q})u_b^{m'}({\mib k} - {\mib q})
 \ , 
\end{eqnarray}
where the repeated indices $l$, $l'$, $m$ and $m'$ are summed over.
By using the approximate expression of the mode functions we obtain the
following expression for the bare response function
$$
\Pi_{ab}^{(0)}(\omega, {\mib q}) =
\int\frac{d^3{\mib k}}{(2\pi)^3}
\frac{1}{4E_a({\mib k})E_b({\mib k} - {\mib q})} 
\cdot
\frac{n_a({\mib k})+n_b({\mib k} - {\mib q})+1}
{\omega-(E_a({\mib k})+E_b({\mib k} - {\mib q}))}  \ . 
$$

\section{Implementation of boundary conditions}

In this section we consider the response of the generalized density matrix
for the external source term $\delta{\cal H}_{\rm ext}$ defined by 
equations
(\ref{23}, \ref{24}) with the boundary conditions already described 
in section 4. 
The interaction term $\delta{\cal H}$ induced by the coupling
between the collective-rotating
condensate and intrinsic meson excitations is still omitted here and
will be taken into account in the next section.
We split 
$\delta{\cal M}$ into $\delta{\cal M}^{(-)}+\delta{\cal M}^{(+)}$
in which $\delta{\cal M}^{(\mp)}$
corresponds to the response for $\delta{\cal H}_{\rm ext}^{(\mp)}\propto
e^{\mp i\omega t}$. Since $\delta{\cal M}^{(\mp)}\propto 
e^{\mp i\omega t}$,
we can easily obtain the solution for (\ref{28}) without $\delta{\cal H}$
as
\begin{equation}\label{31}
\langle a,{\mib k},\tau | \delta{\cal M}^{(\pm)} |b, {\mib k}',\tau' \rangle
=\frac{f_b^{\tau'}({\mib k}')-f_a^\tau({\mib k})}
 {\mp \omega-(E_a^\tau({\mib k})-E_b^{\tau'}({\mib k}'))}
\langle a,{\mib k},\tau | \delta{\cal H}_{\rm ext}^{(\pm)}
|b, {\mib k}',\tau' \rangle ,
\end{equation}
where, from (\ref{29}),
\begin{equation}\label{32}
\langle a,{\mib k},\tau | \delta{\cal H}_{\rm ext}^{(\pm)}
|b, {\mib k}',\tau' \rangle 
=
X_{lm}^{(\pm)}(q)\left[
\pmatrix{u_l^{a*}({\mib k}) \cr -u_l^a({\mib k}) }
\pmatrix{ u_m^{b}({\mib k}') , u_m^{b*}({\mib k}') }
\right]_{\tau\tau'} 
\delta({\mib k}'-{\mib k}\mp{\mib q})e^{\pm i\omega t} \ .
\end{equation}
Here the factor
$\delta({\mib k}'-{\mib k}\mp{\mib q})$ reflects momentum
conservation arising from the matrix element
$\langle {\mib k} |e^{\mp i{\bf q}\cdot{\bf x}} | {\mib k}' \rangle$.

However this particular solution of the linearized 
evolution equations does
not correspond to the desired boundary conditions
$\delta{\cal M}=0$ at time $t=0$. As discussed already 
in section 4 this is adequate to describe
a situation in which a collective rotating 
state is created at time $t$ =0. In order to satisfy this boundary 
condition we add a particular solution of the homogeneous equation
\begin{equation}
i\delta{\dot{\cal M}}=[{\cal H}_0 , \delta{\cal M}]
\end{equation}
to the above-obtained solutions. This can be achieved by turning the phases
$e^{\pm i\omega t}$ into
$e^{\pm i\omega t}-e^{-i(E_a^{\tau}({\bf k})-E_b^{\tau'}({\bf k}'))t}$
in the above expressions (\ref{31}) and (\ref{32}).
The solutions are explicitly given by means of the following formulae. 
For the
matrix elements between two positive energy states $\tau$=+, $\tau'$=+ 
one has
\begin{eqnarray}\label{331}
\langle a,{\mib k},+| \delta{\cal M}^{(\pm)} |b, {\mib k}',+ \rangle
& &=
\frac{n_a({\mib k})-n_b({\mib k}')}{\omega \pm 
(E_a({\mib k})-E_b({\mib k}'
))}
u_l^{a*}({\mib k})X_{lm}^{(\pm)}(q)u_m^{b}({\mib k}') \nonumber\\
& & \ \ \times 2i\sin\frac{\omega\pm (E_a({\mib k})-E_b({\mib k}'))}{2}t
\cdot
\delta({\mib k}'-{\mib k}\mp{\mib q})\nonumber\\
& &\ \ \times
e^{\pm i(\omega\mp(E_a({\bf k})-E_b({\bf k}')))t/2} \ .
\end{eqnarray}
For the
matrix elements between one positive and one
negative energy state $\tau$=+, $\tau'=-$ one has
\begin{eqnarray}\label{332}
\langle a,{\mib k},+| \delta{\cal M}^{(\pm)} |b, {\mib k}',- \rangle
& &=
\frac{n_a({\mib k})+n_b({\mib k}')+1}
{\omega \pm (E_a({\mib k})+E_b({\mib k}'))}
u_l^{a*}({\mib k})X_{lm}^{(\pm)}(q)u_m^{b*}({\mib k}') \nonumber\\
& &\ \ \times 2i\sin\frac{\omega\pm (E_a({\mib k})+E_b({\mib k}'))}{2}t
\cdot
\delta({\mib k}'-{\mib k}\mp{\mib q})\nonumber\\
& &\ \ \times
e^{\pm i(\omega\mp(E_a({\bf k})+E_b({\bf k}')))t/2} \ .
\end{eqnarray}
For the
matrix elements between one negative and one positive
energy state $\tau=-$, $\tau'=+$ one has
\begin{eqnarray}\label{333}
\langle a,{\mib k},-| \delta{\cal M}^{(\pm)} |b, {\mib k}',+ \rangle
& &=
\frac{n_a({\mib k})+n_b({\mib k}')+1}
{\omega \mp (E_a({\mib k})+E_b({\mib k}'))}
u_l^{a}({\mib k})X_{lm}^{(\pm)}(q)u_m^{b}({\mib k}') \nonumber\\
& &\ \ \times 2i\sin\frac{\omega\mp (E_a({\mib k})+E_b({\mib k}'))}{2}t
\cdot
\delta({\mib k}'-{\mib k}\mp{\mib q})\nonumber\\
& &\ \ \times
e^{\pm i(\omega\pm(E_a({\bf k})+E_b({\bf k}')))t/2} \ .
\end{eqnarray}
Finally for the
matrix elements between two negative energy states $\tau=-$, 
$\tau'=-$ 
one has
\begin{eqnarray}\label{334}
\langle a,{\mib k},-| \delta{\cal M}^{(\pm)} |b, {\mib k}',- \rangle
& &=
\frac{n_a({\mib k})-n_b({\mib k}')}{\omega \mp 
(E_a({\mib k})-E_b({\mib k}'
))}
u_l^{a}({\mib k})X_{lm}^{(\pm)}(q)u_m^{b*}({\mib k}') \nonumber\\
& &\ \ \times 2i\sin\frac{\omega\mp (E_a({\mib k})-E_b({\mib k}'))}{2}t
\cdot
\delta({\mib k}'-{\mib k}\mp{\mib q})\nonumber\\
& &\ \ \times
e^{\pm i(\omega\pm(E_a({\bf k})-E_b({\bf k}')))t/2} \ .
\end{eqnarray}
Here the indices $l$ and $m$ are summed over.
At this stage, we are ready to give some comments.
If temperature is equal to 0, then
$\langle a, {\mib k}, \pm | \delta{\cal M} |b, {\mib k}', \pm 
\rangle
=0$ because $n_a({\mib k})=0$. The dominant responses are
$\langle a, {\mib k}, \pm | \delta{\cal M}^{(\mp)} |b, {\mib k}', \mp 
\rangle$
because they have a two-meson pole.
One can understand what physical process these terms represent. 
Let us remember the relation $a_k|n\rangle =\sqrt{n_k}|n-1\rangle$ and
$a_k^{\dagger}|n\rangle = \sqrt{n_k+1}|n+1\rangle$ for boson annihilation
and creation operators and the eigenstates of boson number.
The amplitude of two-meson creation minus two-meson absorption process is 
thus
proportional to $(n_k+1)(n_{k'}+1)-n_k n_{k'}$.
This is nothing but the factor appearing in
$\langle a, {\mib k}, \pm | \delta{\cal M}^{(\mp)} |b, {\mib k}', \mp 
\rangle$.
Thus a net two-meson emission is realized because  
they have the factor $n_a({\mib k})+n_b({\mib k}')+1$.
Let us imagine the relativistic heavy ion
or proton-proton collisions. The restored chiral symmetry spontaneously
breaks down again after rapid cooling.
The spontaneous symmetry break-down
may then be realized at temperature $T\approx 0$
with the explicit chiral symmetry breaking term.
If so-called disoriented chiral condensed state is developed, this state
should be finally relaxed to the normal condensed state due to meson
emissions. In those situations, the response
$\langle a, {\mib k}, \pm | \delta{\cal M}^{(\mp)} |b, {\mib k}', \mp 
\rangle$
will play an important role through two-meson emission from collective 
isospin
rotating mean field (condensate)
and will dominantly contribute to the damping time of the collective
rotation in isospin space. We will return to this consideration
when we estimate the damping time.

\section{Effect of interaction terms}

In this section we take into account of the interaction term $\delta
{\cal H}$
which was neglected in the previous section.
When this term is present the external field can induce a change
$\delta\Gamma$ in (\ref{23}) in the mean
field given by the following formulae
\begin{equation} \label{38}
\delta\Gamma_{ab}
=  \frac{\lambda}{6} \delta_{ab} \sum_{c=1}^{N}\alpha_{cc}
+\frac{\lambda}{3}\alpha_{ab} \ ,
\end{equation}
where the quantity $\alpha$ satisfies the following equation
\begin{equation} \label{43}
\alpha_{ab}=
(X_{ab} +\delta \Gamma_{ab}) \Pi_{ab}^{(0)} \ .
\end{equation}
Let us introduce the dressed response function $\Pi_{ik}$  defined by
\begin{equation} \label{responsea}
\Pi_{ik}(\omega, {\mib q})=
\frac{ \Pi_{ik}^{(0)}(\omega, {\mib q})}
{1-\frac{\lambda}{2} \Pi_{ik}^{(0)}(\omega, {\mib q})},
\end{equation}
for $i=k$ while
\begin{equation} \label{responseb}
\Pi_{ik}(\omega, {\mib q})=
\frac{\Pi_{ik}^{(0)}(\omega, {\mib q})}
{1-\frac{\lambda}{3} \Pi_{ik}^{(0)}(\omega, {\mib q})},
\end{equation}
for $i \ne k$. With this notation
we can write $\alpha$ in the following form:
\begin{eqnarray}\label{44}
& &\alpha_{22}
=\left( X_{22}+X_{11} \frac{\lambda}{6}
\Pi_{11} \right)
\cdot
\Pi_{22}
\cdot\frac{1}
{1-\frac{\lambda}{6} \Pi_{22}\cdot\frac{\lambda}{6} \Pi_{11}},
\nonumber\\
& &\alpha_{11}
={\hbox{\rm (1$\leftrightarrow$ 2 in the above expression)}} \ , 
\nonumber\\
& &\alpha_{12}=X_{12} \Pi_{12} \ ,
\nonumber\\
& &\alpha_{21}=X_{21}
\Pi_{21} \ ,
\nonumber\\
& &\alpha_{ij}=0 \qquad {\hbox{\rm (for $i,j \geq 3$)}} \ .
\end{eqnarray}
The expressions for the physically important functions are thus obtained
with a boundary condition $\delta{\cal M}=0$ at $t=0$ :
\begin{eqnarray}\label{45}
\langle a,{\mib k},\mp |\delta{\cal M}^{(\pm)} |b,{\mib k}',\pm \rangle
&\approx&
\frac{i}{\sqrt{E_a({\mib k})E_b({\mib k}\pm{\mib q})}}\cdot
\frac{n_a({\mib k})+n_b({\mib k}\pm{\mib q})+1}
{\omega-(E_a({\mib k})+E_b({\mib k}\pm{\mib q}))} \nonumber\\
&\times& (X_{ab}^{(\pm)}+\delta\Gamma_{ab}^{(\pm)})
\sin\left(\frac{\omega-(E_a({\mib k})+E_b({\mib k}\pm{\mib q}))}{2}t\right)
\nonumber\\
&\times&
\delta({\mib k}'-{\mib k}\mp{\mib q})
e^{\pm i(\omega+E_a({\bf k})+E_b({\bf k}\pm{\bf q}))t/2} \ .
\end{eqnarray}
Here $\delta\Gamma_{ab}^{(\pm)}(q)$ is related to the amplitudes 
$X^{(\pm)}$
defining the external field via equations
(\ref{38}) and (\ref{44}).

\section{Damping time}
In this section we give the damping time of the collective rotating
condensate in isospin space (or disoriented chiral condensation).
We estimate the damping time
by means of the ratio of the energy density of collective isospin-rotating
condensate to the energy density of emitted mesons per unit time.
To obtain the damping time,
we need the energy density of meson fields originated from the ``external
source" term.

The small change of the generalized density matrix is written by using a
symplectic operator $e^{-i\chi}$ as
\begin{eqnarray}\label{46}
& &{\cal M}=e^{i\chi} {\cal M}_0 e^{-i\chi}
={\cal M}_0+\delta{\cal M}+\delta{\cal M}_2+\cdots \ , \\
& &\ \ \delta{\cal M}=i[\chi , {\cal M}_0 ] \ , \quad
\delta{\cal M}_2=-\frac{1}{2}[ \chi , [\chi , {\cal M}_0]] \ .\nonumber
\end{eqnarray}
Since we have known $\delta{\cal M}$ in the previous section, we can get
$\chi$ as
\begin{equation}\label{47}
\langle a, {\mib k}, \tau | \chi |b, {\mib k}', \tau' \rangle
=\frac{i}{f_a^{\tau}({\mib k})-f_b^{\tau'}({\mib k}')}
\langle a, {\mib k}, \tau | \delta{\cal M} |b, {\mib k}', \tau' 
\rangle .
\end{equation}
The variation of the energy comes from $\delta{\cal M}_2$ because
${\rm Trace}({\cal H}_0 [\chi , {\cal M}_0])=0$ due to
$[{\cal H}_0 , {\cal M}_0]=0$ \cite{VAUTHERIN} :
\begin{eqnarray}\label{48}
\Delta E &=& {\rm Trace} ({\cal H}_0 \delta{\cal M}_2) \nonumber\\
&=& -\frac{1}{2}\sum_{a{\bf k}\tau}\sum_{b{\bf k}'\tau'}
\frac{E_a^{\tau}({\mib k})-E_b^{\tau'}({\mib k}')}
{f_a^{\tau}({\mib k})-f_b^{\tau'}({\mib k}')}
\langle a, {\mib k}, \tau | \delta{\cal M} |b, {\mib k}', \tau' \rangle
\langle b, {\mib k}', \tau' | \delta{\cal M} |a, {\mib k}, \tau \rangle \ .
\end{eqnarray}
This energy shift is calculated by using the functions
$\langle a, {\mib k}, \pm | \delta{\cal M}^{(\mp)} |b, {\mib k}', \mp 
\rangle$
because these terms mainly contribute to the energy shift satisfying the 
energy conservation as will be seen later. As a result we obtain
\begin{eqnarray}\label{49}
& &\Delta E_{ab} = -\frac{1}{2}\int d^3{\mib k}\int d^3{\mib k}'
\frac{E_a({\mib k})+E_b({\mib k}')}
{n_a({\mib k})+n_b({\mib k}')+1}
\nonumber\\
& &\qquad\ \ \ \times
(\langle a, {\mib k}, + | \delta{\cal M}^{(-)} |b, {\mib k}', - \rangle
\langle b, {\mib k}', - | \delta{\cal M}^{(+)} |a, {\mib k}, + \rangle
\nonumber\\
& &\qquad \ \ \ +
\langle a, {\mib k}, - | \delta{\cal M}^{(+)} |b, {\mib k}', + \rangle
\langle b, {\mib k}', + | \delta{\cal M}^{(-)} |a, {\mib k}, - \rangle)
, 
\end{eqnarray}
where $\Delta E=\sum_{ab}\Delta E_{ab}$. Here we can calculate the energy
density by noting that the space volume $V$ is expressed as
$V=\int d^3 {\mib x}=(2\pi)^3 \delta^3({\mib k}={\bf 0})$.

The most important process is the two-pion emission corresponding to
$a=b=2$ among these responses.
In this channel, we denote
$X_{22}^{(+)}=X_{22}^{(-)}\equiv X_{22}$ and
$\delta\Gamma_{22}^{(+)}=\delta\Gamma_{22}^{(-)}\equiv \delta\Gamma_{22}$.
The energy density is expressed as
\begin{eqnarray}\label{50}
\Delta{\cal E}_{\pi\pi}&=&\frac{\Delta E_{\pi\pi}}{V}\nonumber\\
&=&\int\frac{d^3{\mib k}}{(2\pi)^3} (n_2({\mib k})+n_2({\mib k}-{\mib
q})+1)
\left[\frac{1}{E_2({\mib k})}
+\frac{1}{E_2({\mib k}-{\mib q})}\right] \nonumber\\
& &\times
(X_{22}(q)+\delta\Gamma_{22}(q))^2
\ \frac{\sin^2 \left[ (\omega-E_2({\mib k})-E_2({\mib k}-{\mib q}))t /2
\right] } { (\omega-E_2({\mib k})-E_2({\mib k}-{\mib q}))^2}\ .
\end{eqnarray}
Using the relation $(\sin(\omega t/2)/(\omega/2))^2 \approx 2\pi
t\delta(\omega)$ for large $t$, which ensures the energy conservation,
we find that the contribution of the two pion excitation to the energy
density grows in proportion to time,
\begin{equation}\label{50a}
\Delta{\cal E}_{\pi\pi} = \gamma_{\pi\pi} (q) t \,
\end{equation}
where
\begin{eqnarray}\label{50b}
\gamma_{\pi\pi} (q) & = &
\int\frac{d^3{\mib k}}{(2\pi)^3}
(n_2({\mib k})+n_2({\mib k}-{\mib q})+1)
\left[\frac{1}{E_2({\mib k})}
+\frac{1}{E_2({\mib k}-{\mib q})}\right] \nonumber\\
& &\times
\frac{1}{4}(X_{22}(q)+\delta\Gamma_{22}(q))^2
2\pi\delta(\omega-E_2({\mib k})-E_2({\mib k}-{\mib q})) \,
\end{eqnarray}
gives the rate of the increase of the energy density associated with
two pion excitation which carries net momentum $q_{\mu} = (\omega, 
{\mib q})$.
Similar expressions $\gamma_{\pi \sigma}$ and $\gamma_{\sigma
\sigma}$ for the $\sigma\pi$- and $\sigma\sigma$-channels are also
obtained.

The damping time $\tau (q)$
of the collective rotating condensate in isospin space
is simply estimated in terms of the energy density of rotating 
condensate and the energy density of emitted meson per unit time :
\begin{equation}\label{53}
\tau (q) = {\cal E}_0 / \gamma (q) \ .
\end{equation}
Here, the numerator is the energy density of the condensate 
originated from the collective rotation in isospin space :
\begin{eqnarray}
{\cal E}_0 
&=& \frac{1}{2}{{\bar \pi}}_a({\mib x})
{{\bar \pi}}_a({\mib x})
+\frac{1}{2}\nabla {\bar \varphi}_a({\mib x}) 
\cdot\nabla {\bar \varphi}_a({\mib x}) \nonumber\\
&=& \frac{1}{2}\varphi_0^2(\omega^2+{\mib q}^2) \ ,
\end{eqnarray}
where we have set the pion mass equal to zero in the lowest order
approximation for $\tau(q)$ according to the results obtained in
Appendix A and we have neglected a correction from the
quantum fluctuations which we had shown to be small (see section
8 of I).

The denominator
$$
\gamma (q) = \gamma_{\pi\pi} (q) + \gamma_{\pi\sigma} (q) +
\gamma_{\sigma\sigma} (q)
$$
is the total rate of the change of the energy density due to two meson
emissions.

\section{Discussion}

We estimate the associated damping time and the meson emission rate
in this section.
Let us consider a simple case with ${\mib q}={\bf 0}$.
The rate of the change of the energy density due to two-pion emission
with three momentum ${\mib q}={\bf 0}$ is simply written as
\begin{eqnarray}\label{51}
\gamma_{\pi\pi} (\omega, {\bf 0}) &=&
\frac{1}{8\pi}\left[X_{22}(\omega,{\bf 0})
\left(1+\frac{\lambda}{2}\Pi_{22}(\omega,{\bf 0})\right)
\right]^2
\sqrt{\omega^2-4M_2^2} \nonumber\\
& &\times \left(2n_2\left(\frac{\omega}{2}\right)+1\right)
\theta(\omega-2M_2) \ ,
\end{eqnarray}
where we neglected $\Pi_{11}(q)$ which corresponds to the
$\sigma\sigma$ excitation because two sigma mesons are too heavy
in our consideration.
For the $\sigma\pi$-channel at $T=0$ we obtain
\begin{eqnarray}\label{52}
& &\gamma_{\pi\sigma} (\omega,{\bf 0} )=
\frac{1}{4\pi}\left[|X_{12}(\omega,{\bf 0})|
\left(1+\frac{\lambda}{3}\Pi_{12}(\omega,{\bf 0})\right)
\right]^2 
2k_0\theta(\omega-M_1-M_2) \ , \\
& &k_0=
\frac{1}{2\omega}\sqrt{(\omega^2-(M_1+M_2)^2)
(\omega^2-(M_1-M_2)^2)} \ . \nonumber
\end{eqnarray}

Let us give an order estimation of the damping time
in the case of ${\mib q}={\bf 0}$.
We numerically estimate the typical damping time at 
$\omega=2\sqrt{2}M_\pi$ where $M_{\pi}(\neq 0)$ is the physical 
pion mass. The reason is that 
the factor $\omega^2/\sqrt{\omega^2-4M_{\pi}^2}$ which appears in
(\ref{53}) with (\ref{51})
has a minimum value if $M_2$ is close to $M_\pi$
due to the isospin-rotating effect.
In this case, the energy density
of collective isospin-rotating condensate is
$f_\pi^2\omega^2/2 \approx (160{\rm MeV})^4$
where we used $\varphi_0=f_\pi=93$
MeV (the pion decay constant).
Note that the change in the potential energy density induced by the chiral
symmetry
breaking term is $2M_{\pi}^2 f_{\pi}^2$ where we used 
$c=f_{\pi}M_{\pi}^2$.
The validity of our approximation scheme requires $f_{\pi}^2\omega^2/2 \gg
2M_{\pi}^2f_{\pi}^2$, i.e., $\omega \gg 2M_{\pi}$, which is reasonably
satisfied for our choice of $\omega$.
The damping time is roughly estimated as
\begin{equation}\label{54}
\tau=\tau_{\pi\pi}\approx 40 \ {\rm fm}/c \quad
{\hbox{\rm at $\omega=2\sqrt{2}M_\pi$ and ${\mib q}={\bf 0}$}} \ ,
\end{equation}
where we used $m_\pi\approx M_2\approx 0$, $M_\pi=140$ MeV,
$m_\sigma \approx M_1 \approx M_\sigma=500$ MeV and $\lambda=86.7$ and
 we introduced the three momentum cutoff $\Lambda=1$ GeV\cite{TSUE}
 to calculate the response function $\Pi_{22}$. This is rather
long compared with the collision time of heavy ion collision and/or
the formation time of so-called disoriented chiral condensation
about a few fm/$c$. When we take into account of the small finite
${\mib q}$, the damping time becomes slightly longer.  

It is
interesting to note that the calculation we have just made also
provides information on the meson emission rate. Indeed, from
equation (\ref{54}) we find that the energy density decrease per
unit time is of the order of (100 MeV)$^4$ per fm/$c$. If we assume
that the classical field configuration occupies a volume of the
order of 1000 cubic fm this implies an emission rate of 100 mesons
per fm/$c$.  

Further studies of the present model are worth
being pursued. For instance it would be interesting to consider
initial conditions for which there would be a coupling between
collective rotations and small intrinsic oscillations. Comparing the
damping times of rotations and vibrations separately would indeed
teach us which of the two collective modes is selected by
dissipation effects. Another interesting problem is to combine the
effect of the expansion of the meson field, already considered in
references \cite{SERREAU} and \cite{BRAGHIN}, with collective
isospin rotations.

\acknowledgements

Two of the authors (D. V. and Y. T.) would like to thank
Prof. D. Boyanovsky and Prof. H. J. de Vega for valuable discussions
and comments. One of us (D. V.) wishes to express his appreciation
to the Japan Society for the Promotion of Science for support to
visit the Yukawa Institute where this work was started.  Y. T. is
grateful to the members of the Laboratoire de Physique Th\'eorique
des Particules El\'ementaires (LPTPE) of Universit\'e Pierre et
Marie Curie for their warm hospitality. He also would like to
express his sincere thanks to Nishina Memorial Foundation for
financial support to visit the LPTPE.  T. M. is grateful to the
Laboratoire de Physique Th\'eorique des Particules El\'ementaires
(LPTPE) of Universit\'e Pierre et Marie Curie for the support and
the hospitality which he received during his visit.

\appendix

\section{Pionic excitation in the chiral limit}
In this appendix, we calculate the pionic response to a weak
external perturbation in the linearized approximation.  It is shown
here that we can obtain in this framework the pion mass which is
consistent with the Goldstone theorem requiring that the pion
mass is exactly equal to zero in the chiral limit. We introduce a
source term $J_a (x)$ acting on the second component of the
$N$-component scalar field with O(N)-symmetry.  In general we can
extract the value of the square mass of the meson field from the
functional $\delta \langle \varphi (x) \rangle / \delta J(y)$ which is
the inverse propagator in the limit of the small source.  Our task is
thus to calculate the mean field value $\langle\varphi\rangle$
depending on $J$.

The equation of motion for the mean field $\varphi$ with a source 
term $J$ is obtained as
\begin{equation} \label{pion1}
\left( \Box+ m^2 -\frac{\lambda}{3} \varphi \times \varphi  
\right) 
{\varphi} = J  \mbox{,}
\end{equation}
where $m_{ab}^2$ has the same form as (\ref{gap}) by denoting 
${\bar \varphi}$ as $\varphi$. 
If $J=0$, then the above equation is reduced to (\ref{MEANFIELD}) 
with (\ref{gap}) in section 2. We denote the Feynman propagator 
$S(x,x)$ appearing in $m^2(x)$ as $S^{(0)}(x,x)$ 
for the case $J(x)=0$ in this appendix. 
It is here noted that the propagator $S^{(0)}$ 
is diagonal in the isospin quantum number. 

Let us take $J_a(x)$ to be of the form
\begin{equation}\label{pion4}
J_a(x)=\epsilon \delta_{a2} \exp (i\omega t-i {\mib q}\cdot{\mib x}) \ .
\end{equation}
The propagator $S(x,x)$ is changed according to the change of 
$m^2(x)$, $m^2(x)\rightarrow m^2(x)+\delta m^2(x)$, 
from the case $J(x)=0$ to $J(x)\neq 0$, namely, 
$S(x,x)=S^{(0)}(x,x)+\delta S(x,x)$. 
We write the change $m^2(x)$ as 
\begin{equation}\label{pion5}
\delta m_{ab}^2(x) = \delta m_{ab}^2 
\exp (i\omega t-i{\mib q}\cdot{\mib x}) \ .
\end{equation}
As a result, we obtain $\delta S(x,x)$ :
\begin{equation}\label{pion6}
\delta S_{ab}(x,x)=\delta m_{ab}^2 \cdot \Pi_{ab}^{({\rm pol})}(q^2) 
\exp(i\omega t-i{\mib q}\cdot {\mib x}) \ ,
\end{equation}
where $\Pi_{ab}^{({\rm pol})}(q^2)$ is the polarization tensor which is 
expressed in terms of the Fourier transform of $S^{(0)}(x,x)$ being
$S^{(0)}(p)$ as 
\begin{equation}\label{pion7}
\Pi_{ab}^{({\rm pol})}(q^2)=\int \frac{d^4 p}{(2\pi)^4} S_{aa}^{(0)}(p+q)
S_{bb}^{(0)}(p) \ .
\end{equation}

For the source term (\ref{pion4}) we use the following Ansatz for 
$\delta m^2$ : 
\begin{equation}\label{pion8}
\delta m^2=\pmatrix{
                    0 & \eta & 0 & \cdots \cr
                    \eta & 0 & 0 & \cdots \cr
                    0 & 0 & 0 & \cdots \cr
                    \vdots & \vdots & \vdots & \ddots
                    } \ .
\end{equation}
An ansatz of the same form in the matrix structure 
is assumed for the quantity 
$\delta S(x,x)$. Then the linearized evolution equation 
for the small shift $\delta\varphi(x)\equiv 
\varphi(x)-\varphi^{(0)}(x)$ reads 
\begin{equation}\label{pion9}
\left( -q^2+m^2-\frac{\lambda}{3} \varphi^{(0)} \times
\varphi^{(0)}  \right) \delta\varphi
+\delta m^2 \varphi^{(0)} 
 -\frac{\lambda}{3}
( \delta \varphi \times \varphi^{(0)}
+ \varphi^{(0)} \times \delta\varphi) \varphi^{(0)}
= J \ ,
\end{equation}
where $\varphi^{(0)}(x)$ is the value of the mean field in the case 
$J(x)=0$. 
For the mass matrix we obtain 
\begin{equation}\label{pion10}
\delta m^2=\frac{\lambda}{3}
(\delta\varphi\times\varphi^{(0)}
+\varphi^{(0)}\times\delta\varphi)
+\frac{\lambda}{3}\delta S(x,x) \ 
\end{equation}
which is consistent with the ansatz (\ref{pion8}) because 
$\varphi_a^{(0)}(x)=\delta_{a1}\varphi_0$ and the second component 
of $\delta\varphi(x)$ is only non-zero according to 
the source (\ref{pion4}).
It should be noted that for the source (\ref{pion4}) there is 
no change in Trace $\delta S(x,x)$ and no change in $\varphi^2(x)$ 
at least in the lowest order. By using the result (\ref{pion6}) and 
equation (\ref{pion10}), we obtain
\begin{equation}\label{pion11}
\delta m^2 \varphi^{(0)} = \frac{\lambda}{3}\varphi_0^2
\frac{1}{1-\frac{\lambda}{3}\Pi_{12}^{({\rm pol})}}
\delta\varphi \ ,
\end{equation}
where we used $\varphi^{(0)} \cdot \delta\varphi = 0$. 
From the linearized evolution equation (\ref{pion9}) and the 
above relation (\ref{pion11}) it is possible to express the change 
$\delta\varphi$ in terms of the condensate 
in the presence of the external source $J$. 
It is given by 
\begin{equation}\label{pion12}
\left(-q^2+m^2+\frac{\lambda}{3}\varphi_0^2\cdot\frac{\lambda}{3}
\Pi_{12}^{({\rm pol})}\cdot
\frac{1}{1-\frac{\lambda}{3}\Pi_{12}^{({\rm pol})}}\right)
\delta\varphi = J\ .
\end{equation}
The left-hand side is related to the dressed propagator. This provides a
prescription to calculate the pion mass. 
Indeed from the above equation (\ref{pion12}) 
we obtain the $J$-dependent part of the mean field $\varphi(x)$, 
that is, the shift $\delta\varphi(x)$. This gives the 
inverse propagator for the meson field 
which is precisely the functional derivative 
$\delta\varphi(x)/ \delta J(y)$.
The square of the pion mass 
$M_{\pi}^2$ is therefore just the value of the bracket calculated 
at $q=0$, i.e., 
\begin{equation}\label{pion13}
M_{\pi}^2=\mu^2+\frac{\lambda}{3}\varphi_0^2
\cdot\frac{\lambda}{3}\Pi_{12}^{({\rm pol})}
\cdot\frac{1}{1-\frac{\lambda}{3}\Pi_{12}^{({\rm pol})}}
\ , 
\end{equation}
where $\Pi_{12}^{({\rm pol})}$ means $\Pi_{12}^{({\rm pol})}(q^2=0)$. 
Here, we have used 
the fact that the mass matrix has the following matrix structure 
\cite{PLB} :
\begin{equation}\label{pion14}
m^2=\pmatrix{
             M^2 & 0 & 0 & 0 & \cdots \cr
             0 & \mu^2 & 0 & 0 & \cdots \cr
             0 & 0 & \mu^2 & 0 & \cdots \cr
             0 & 0 & 0 & \mu^2 & \cr
             \vdots & \vdots & \vdots & & \ddots
} \ , 
\end{equation}
where the masses $M$ and $\mu$ satisfy the self-consistency conditions :
\begin{eqnarray}\label{pion15}
& & M^2 = m_0^2+\frac{\lambda}{2}\varphi_0^2
 +\frac{\lambda}{2}G(M^2) +(N-1)\frac{\lambda}{6}G(\mu^2)\ , \nonumber\\
& & \mu^2 = m_0^2+\frac{\lambda}{6}\varphi_0^2
 +\frac{\lambda}{6}G(M^2)+(N+1)\frac{\lambda}{6}G(\mu^2)  
\end{eqnarray}
as well as the condensate minimization equation 
\begin{equation}\label{pion16}
M^2=\frac{\lambda}{3}\varphi_0^2 \ .
\end{equation}
The function $G(m^2)$ is defined by
\begin{equation}\label{pion17}
G(m^2)=-\int\frac{d^4p}{(2\pi)^4} \frac{i}{p^2-m^2+i\varepsilon} 
\ . 
\end{equation}
One can further use at this point the expression of the polarization 
tensor (\ref{pion7}) in terms of $G(m^2)$, 
that is, $
\Pi_{12}^{({\rm pol})}(q=0)=[G(M^2)-G(\mu^2)]/(M^2-\mu^2) 
$.
Also, $\mu^2$ can be expressed in terms of $G(m^2)$. 
Namely, by subtracting the second equation from the first one 
in (\ref{pion15}) and by using the minimization equation 
(\ref{pion16}), we obtain 
$\mu^2=-\lambda[G(M^2)-G(\mu^2)]/3$.
Finally, from (\ref{pion13}), we can derive the following relation 
by using the above-derived expressions of the polarization tensor 
and $\mu^2$ :
\begin{equation}\label{pion19}
\left(1-\frac{\lambda}{3}\Pi_{12}^{({\rm pol})}(q=0)\right)M_{\pi}^2=0 \ .
\end{equation}
This gives $M_{\pi}=0$ which is consistent with the Goldstone 
theorem.

\end{document}